\documentstyle[preprint,prc,aps,epsfig]{revtex}
\tighten
\begin{document}
\draft
\title{Antikaon condensation and the metastability of protoneutron stars}
\author{Sarmistha Banik and Debades Bandyopadhyay}
\address{Saha Institute of Nuclear Physics, 1/AF Bidhannagar, 
Calcutta 700 064, India}

\maketitle

\begin{abstract}
We investigate the condensation of $\bar K^0$ meson along with  
$K^-$ condensation in the neutrino trapped matter with and without 
hyperons. Calculations are performed in the 
relativistic mean field models in which both the baryon-baryon
and (anti)kaon-baryon interactions are mediated by meson exchange. 
In the neutrino trapped matter relevant to protoneutron stars, the critical
density of $K^-$ condensation is shifted considerably to higher density
whereas that of $\bar K^0$ condensation is shifted slightly to higher density 
with respect to that of the neutrino free case. The onset of $K^-$ condensation
always occurs earlier than that of $\bar K^0$ condensation. A significant 
region of maximum mass protoneutron stars is found to contain $\bar K^0$ 
condensate for larger values of the antikaon potential. With the appearance 
of $\bar K^0$ condensation, there is a region of symmetric nuclear matter 
in the inner core of a protoneutron star. It is found that the maximum mass 
of a protoneutron star containing $K^-$ and $\bar K^0$ condensate is greater 
than that of the corresponding neutron star. We revisit the implication of this
scenario in the context of the metastability of protoneutron stars and their 
evolution to low mass black holes.  

\end{abstract}
\newpage
\section{Introduction}
It was suggested by Woosley et al. \cite{Woo} that progenitor stars
heavier than $\sim 25 M_{\odot}$ would collapse into black holes. In this
scenario, stars first explode; exhibit light curves of type II supernova
and return matter to the galaxy before going into black holes. This
issue got impetus after the explosion of SN 1987A. One of the revealing 
features of SN 1987A is that a neutrino was observed in Kamiokande II at the 
twelveth second. So far there has been no observation of a pulsar within it.
Moreover, the light curve fades away leading to the speculation that the
compact object in SN 1987A has collapsed into a low mass black hole. If this
picture of low mass black hole formation is true for SN 1987A, the newly born
hot and neutrino rich star, called protoneutron star, was stable over twelve
seconds or more before collapsing into a black hole.

In recent years, there have been several works by various groups 
\cite{Bro2,Brown,Pra97} to understand what is the mechanism behind the 
stability of protoneutron stars for short times. 
In the "conventional" scenario where (proto)neutron stars are made up of 
nucleons and leptons, the protoneutron star has a slightly smaller 
maximum mass than that of the neutron star. The "window" of maximum masses is
very small in this case. However, the scenario dramatically changes with the
formation of $K^-$ condensation in dense matter as found in previous 
calculations
\cite{Bro2,Brown,Pra97}. They showed that leptons could stabilize much larger 
maximum mass for a protoneutron star in the presence
of $K^-$ condensation during the evolution. In the following paragraph, we
briefly review the previous calculations of $K^-$ condensation in dense matter
relevant to (proto)neutron stars.

With the pioneering work of Kaplan and Nelson \cite{Kap}, a considerable 
interest has been generated in the study of antikaon ($\bar K$) condensation
in dense matter in recent years. In a chiral $SU(3)_L\times SU(3)_R$ model,
baryons directly couples with (anti)kaons. 
The effective mass $m_K^*$ of antikaons decreases with density because of 
the strongly attractive $K^-$-baryon interaction in dense matter.
Consequently, the in-medium energy ($\omega_{K^-}$) of $K^-$ meson in the zero 
momentum state also decreases with density. The $s$-wave $K^-$ condensation 
sets in when $\omega_{K^-}$ equals to the chemical potential of $K^-$ meson.
Later, this chiral model was adopted by other groups to study $K^-$ 
condensation in the core of neutron stars \cite{Bro92,Tho,Ell} using 
kaon-nucleon scattering data \cite{Bro94,Li} and $K^-$ atomic data \cite{Lee}.
On the other hand, $K^-$ condensation has been studied in a different kind of
model which is an extension of the Walecka model \cite{Mut,Kno,Sch,Gle99}.
In this model, kaons interact with baryons through the exchange of mesons. 
It is found that the threshold density of $K^-$ condensation in various
calculations depends on the equation of state and parameters, in particular
on the antikaon optical potential.
The net effect of $K^-$ condensation in neutron star matter is that $K^-$
condensate replaces electrons in maintaining charge neutrality and softens 
the equation of state. Due to the softening of the equation of state,
the masses of the stars are reduced in the presence of $K^-$ condensate
\cite{Tho,Mut,Kno,Sch,Gle99}. It was also found that 
in the presence of hyperons,
$K^-$ condensation was delayed to higher density and might not even exist 
in the maximum mass stars. Protoneutron stars
with $K^-$ condensate were studied by various groups \cite{Bro2,Pra97,Tho,Pon}
and shown to have maximum masses larger than those of cold neutron stars - a
reversal from the "conventional scenario". The theoretical studies   
based on the above mentioned models \cite{Bro94,Kno,Sch} and also on 
Nambu$-$Jona-Lasinio model \cite{Lut}, yield a repulsive optical potential for 
$K^+$ in nuclear medium. Therefore, $K^+$ condensation may not be a 
possibility in (proto)neutron stars.

In a recent calculation \cite{Pal}, the formation of 
$\bar K^0$-meson condensation in neutron stars has been investigated 
within a relativistic mean field approach \cite{Ser} where the interaction
between the baryons and antikaons are generated by the exchange of $\sigma$,
$\omega$, and $\rho$ mesons. It is found that the $\rho$-meson field is
repulsive for $K^-$ meson, whereas it is attractive for $\bar K^0$ meson 
which is an isodoublet partner of $K^-$ meson. Consequently, the in-medium
energy ($\omega_{\bar K^0}$) of $\bar K^0$ meson is lowered compared with
that of $K^-$ meson thereby making $\bar K^0$ meson condensation more
favorable in neutron star matter. The critical density for $s$-wave neutral 
$\bar K^0$ meson condensation is governed by the condition
$\omega_{\bar K^0} = 0$. It was found that the critical densities for 
$K^-$ and $\bar K^0$ condensation depended sensitively on the choice of the 
antikaon optical potential depth and more strongly on the nuclear equation of 
state (EOS). The threshold density of $\bar K^0$ condensation
always lie above that of $K^-$ condensation. With the appearance of $K^-$ and 
$\bar K^0$ condensate, the overall equation of state becomes softer than
the situation without antikaon condensation leading to a reduction in 
the maximum masses of neutron stars.  With the onset of only 
$K^-$ condensation, the proton fraction rises dramatically and 
even crosses the neutron fraction at some density because 
of charge neutrality. 
With the onset of $\bar K^0$ condensation, there is a competition in the
formation of ${K^-}-p$ and ${\bar K^0}-n$ pairs resulting in a perfectly
symmetric matter for nucleons and antikaons inside neutron stars. In the
presence of hyperons, it was found that the formation of antikaon condensation
was delayed to higher densities and the maximum mass neutron star contained
$\bar K^0$ condensate for larger values of the antikaon potential depth 
\cite{Pal}.  

So far, there is no calculation of $\bar K^0$ condensation and its 
impact on the gross properties of protoneutron stars. 
In this paper, we investigate the effect of antikaon condensation with    
emphasis on the role of $\bar K^0$ condensate to determine the
composition and structure of protoneutron stars in the standard 
meson exchange model \cite{Ser}. In this calculation, we adopt the
usual relativistic mean field Lagrangian \cite{Sch,Gle99,Pal} for baryons 
interacting via meson exchanges. Also, we include the self-interaction of the 
scalar meson and the nonlinear $\omega$ meson term in the calculation 
\cite{Sch}. (Anti)kaon-baryon interaction is treated on the same footing as 
the baryon-baryon interaction. The Lagrangian density for (anti)kaons is taken 
from Ref. \cite{Gle99,Pal}. We shall show within this model that
$\bar K^0$ condensate may also exist inside a protoneutron star and has a
significant influence on the star's properties and evolution. 

The paper is organized as follows. In section II we describe the 
relativistic mean field (RMF) model of strong interactions. The relevant 
equations for (proto)neutron star matter with antikaon condensates are 
summarized 
in this model. In section III the parameters of the model are discussed and 
results of antikaon condensates in (proto)neutron star matter are presented. 
Section IV is devoted to the summary and conclusions.

\section{The Formalism}
We describe the charge neutral and beta-equilibrated matter consisting of
baryons, electrons, muons and electron type neutrinos in the presence
of antikaon condensates. The starting point in the present approach 
is a relativistic field theoretical model of baryons and (anti)kaons 
interacting by the exchange of scalar $\sigma$, isoscalar vector $\omega$, and 
vector isovector $\rho$ mesons and
two additional hidden-strangeness mesons, the scalar meson $f_0$(975)
(denoted hereafter as $\sigma^*$) and the vector meson $\phi$(1020) to allow
for hyperon-hyperon interaction \cite{Sch,Sch94}. The
total Lagrangian density consists of the baryonic, kaonic and 
leptonic parts, i.e. ${\cal L} = {\cal L}_B + {\cal L}_K + {\cal L}_l$,
Here we consider all the species of the baryon octet 
$B\equiv \{n,p,\Lambda,\Sigma^+,\Sigma^-,\Sigma^0,\Xi^-,\Xi^0 \}$.
The baryonic Lagrangian density is given by
\begin{eqnarray}
{\cal L}_B &=& \sum_B \bar\psi_{B}\left(i\gamma_\mu{\partial^\mu} - m_B
+ g_{\sigma B} \sigma - g_{\omega B} \gamma_\mu \omega^\mu 
- \frac{1}{2} g_{\rho B} 
\gamma_\mu{\mbox{\boldmath $\tau$}}_B \cdot 
{\mbox{\boldmath $\rho$}}^\mu \right)\psi_B\nonumber\\
&& + \frac{1}{2}\left( \partial_\mu \sigma\partial^\mu \sigma
- m_\sigma^2 \sigma^2\right) - U(\sigma) \nonumber\\
&& -\frac{1}{4} \omega_{\mu\nu}\omega^{\mu\nu}
+\frac{1}{2}m_\omega^2 \omega_\mu \omega^\mu
- \frac{1}{4}{\mbox {\boldmath $\rho$}}_{\mu\nu} \cdot
{\mbox {\boldmath $\rho$}}^{\mu\nu}
+ \frac{1}{2}m_\rho^2 {\mbox {\boldmath $\rho$}}_\mu \cdot
{\mbox {\boldmath $\rho$}}^\mu  + {\cal L}_{YY}~.
\end{eqnarray}
Here $\psi_B$ denotes the Dirac spinor for baryon B with vacuum mass $m_B$
and isospin operator ${\mbox {\boldmath $\tau$}}_B$. The scalar 
self-interaction term \cite{Bog} is, 
\begin{equation}
U(\sigma) = \frac{1}{3} g_2 \sigma^3 + \frac{1}{4} g_3 \sigma^4 ~.
\end{equation}
The Lagrangian density (${\cal L}_{YY}$)
responsible for hyperon-hyperon interaction is given by,
\begin{eqnarray}
{\cal L}_{YY} &=& \sum_B \bar\psi_{B}\left(
g_{\sigma^* B} \sigma^* - g_{\phi B} \gamma_\mu \phi^\mu 
\right)\psi_B\nonumber\\
&& + \frac{1}{2}\left( \partial_\mu \sigma^*\partial^\mu \sigma^*
- m_{\sigma^*}^2 \sigma^{*2}\right) 
-\frac{1}{4} \phi_{\mu\nu}\phi^{\mu\nu}
+\frac{1}{2}m_\phi^2 \phi_\mu \phi^\mu~.
\end{eqnarray}

The Lagrangian density for (anti)kaons in the minimal coupling scheme is given
by \cite{Gle99}
\begin{equation}
{\cal L}_K = D^*_\mu{\bar K} D^\mu K - m_K^{* 2} {\bar K} K ~,
\end{equation}
where the covariant derivative 
$D_\mu = \partial_\mu + ig_{\omega K}{\omega_\mu} + ig_{\phi K}{\phi_\mu} 
+ i g_{\rho K} 
{\mbox{\boldmath $\tau$}}_K \cdot {\mbox{\boldmath $\rho$}}_\mu$. 
The isospin doublet for kaons
is denoted by $K\equiv (K^+, K^0)$ and that for antikaons is  
$\bar K\equiv (K^-, \bar K^0)$. The effective mass of (anti)kaons in this
minimal coupling scheme is given by
\begin{equation}
m_K^* = m_K - g_{\sigma K} \sigma - g_{\sigma^* K} \sigma^* ~,
\end{equation}
where $m_K$ is the bare kaon mass. In the mean field approximation (MFA)
\cite{Ser} adopted here, the meson fields are replaced by their 
expectation values. Only the time-like components
of the vector fields, and the isospin 3-component of $\rho$-meson 
field have non-vanishing values in a uniform and static matter.
The mean meson fields are denoted by $\sigma$, $\sigma^*$, $\omega_0$, 
$\phi_0$ and $\rho_{03}$.

The dispersion relation representing the in-medium energies of 
$\bar K\equiv (K^-, \bar K^0)$ for $s$-wave (${\bf k}=0$) condensation 
is given by
\begin{equation}
\omega_{K^-,\: \bar K^0} = m_K^* - g_{\omega K} \omega_0 - g_{\phi K} \phi_0  
\mp \frac{1}{2} g_{\rho K} \rho_{03} ~,
\end{equation}
where the isospin projection $I_{3\bar K} =\mp 1/2$ for the mesons 
$K^-$ ($-$ sign) and $\bar K^0$ (+ sign) are explicitly written in the
expression. Since the $\sigma$ and $\omega$ fields generally increase with 
density  and both the terms containing $\sigma$ and $\omega$ fields in Eq.(6) 
are attractive for antikaons, the in-medium energies of 
$\bar K$ decrease in nuclear medium. On the other hand, in nucleon-only matter
$\rho_{03} \equiv n_p - n_n$ ($n_p$ and $n_n$ are the proton and 
neutron densities) is negative; thus the $\rho$-meson field favors the 
formation of $\bar K^0$ condensation over that of $K^-$ condensation. In 
hyperon matter, the repulsive $\phi$ meson term may delay the onset of 
antikaon condensation \cite{Sch}.  
The in-medium energies of kaons $K\equiv (K^+, K^0)$ are given by, 
\begin{equation}
\omega_{K^+,\: K^0} = m_K^* + g_{\omega K} \omega_0 + g_{\phi K} \phi_0 
\pm \frac{1}{2} g_{\rho K} \rho_{03} ~.
\end{equation}
It is to be noted here that kaon condensation may be impossible in the neutron 
star matter because the $\omega$-meson term is repulsive for kaons and dominates
over the attractive $\sigma$-meson term at higher densities.
However, the attractive $\phi$ meson term may decrease kaon energies 
in the presence of hyperons \cite{Sch}.

The meson field equations in the presence of baryons and antikaon condensates 
are derived from Eqs. (1)-(4) as
\begin{eqnarray}
m_\sigma^2\sigma &=& -\frac{\partial U}{\partial\sigma}
+ \sum_B g_{\sigma B} n_B^S 
+ g_{\sigma K} \sum_{\bar K} n_{\bar K} ~,\\ 
m_{\sigma^*}^2\sigma^* &=& \sum_B g_{\sigma^* B} n_B^S 
+ g_{\sigma^* K} \sum_{\bar K} n_{\bar K} ~,\\ 
m_\omega^2\omega_0 &=& \sum_B g_{\omega B} n_B
- g_{\omega K} \sum_{\bar K} n_{\bar K} ~,\\ 
m_\phi^2\phi_0 &=& \sum_B g_{\phi B} n_B
- g_{\phi K} \sum_{\bar K} n_{\bar K} ~,\\ 
m_\rho^2\rho_{03} &=& \sum_B g_{\rho B} I_{3B} n_B 
+ g_{\rho K} \sum_{\bar K} I_{3\bar K} n_{\bar K} ~.
\end{eqnarray}
Here the scalar and number density of baryon $B$ are respectively
\begin{eqnarray}
n_B^S &=& \frac{2J_B+1}{2\pi^2} \int_0^{k_{F_B}} 
\frac{m_B^*}{(k^2 + m_B^{* 2})^{1/2}} k^2 \ dk ~,\\
n_B &=& (2J_B+1)\frac{k^3_{F_B}}{6\pi^2} ~, 
\end{eqnarray}
with effective baryonic mass $m_B^*=m_B - g_{\sigma B}\sigma 
- g_{\sigma^* B}\sigma^*$,
Fermi momentum $k_{F_B}$, spin $J_B$, and isospin projection
$I_{3B}$. Note that for $s$-wave ${\bar K}$ condensation, the scalar and
vector densities of antikaons are same and those are given 
by \cite{Gle99}
\begin{equation}
n_{K^-,\: \bar K^0} = 2\left( \omega_{K^-, \bar K^0} + g_{\omega K} \omega_0 
+ g_{\phi K} \phi_0 \pm \frac{1}{2} g_{\rho K} \rho_{03} \right) {\bar K} K  
= 2m^*_K {\bar K} K  ~.
\end{equation}
The total energy density $\varepsilon = \varepsilon_B + \varepsilon_l
+ \varepsilon_{\bar K}$
has contributions from baryons, leptons, and antikaons. The baryonic 
plus leptonic energy density is
\begin{eqnarray}
{\varepsilon_B} + {\varepsilon_l} &=& \frac{1}{2}m_\sigma^2 \sigma^2 
+ \frac{1}{3} g_2 \sigma^3 
+ \frac{1}{4} g_3 \sigma^4  + \frac{1}{2}m_{\sigma^*}^2 \sigma^{*2} 
+ \frac{1}{2} m_\omega^2 \omega_0^2 + \frac{1}{2} m_\phi^2 \phi_0^2 
+ \frac{1}{2} m_\rho^2 \rho_{03}^2  \nonumber \\
&& + \sum_B \frac{2J_B+1}{2\pi^2} 
\int_0^{k_{F_B}} (k^2+m^{* 2}_B)^{1/2} k^2 \ dk
+ \sum_l \frac{1}{\pi^2} \int_0^{K_{F_l}} (k^2+m^2_l)^{1/2} k^2 \ dk 
+ \frac{\mu_{\nu_e}^4}{8\pi^2} ~,
\end{eqnarray}
where $l$ goes over electrons and muons.
The last term corresponds to the energy density of neutrinos as required in a
protoneutron star matter. The energy density for antikaons is
\begin{equation}
\varepsilon_{\bar K} = m^*_K \left( n_{K^-} + n_{\bar K^0} \right) .
\end{equation}
Since antikaons form $s$-wave Bose condensates, they do not directly
contribute to the pressure so that the pressure is due to baryons and 
leptons only
\begin{eqnarray}
P &=& - \frac{1}{2}m_\sigma^2 \sigma^2 - \frac{1}{3} g_2 \sigma^3 
- \frac{1}{4} g_3 \sigma^4  - \frac{1}{2}m_{\sigma^*}^2 \sigma^{*2} 
+ \frac{1}{2} m_\omega^2 \omega_0^2 + \frac{1}{2} m_\phi^2 \phi_0^2 
+ \frac{1}{2} m_\rho^2 \rho_{03}^2 \nonumber \\
&& + \frac{1}{3}\sum_B \frac{2J_B+1}{2\pi^2} 
\int_0^{k_{F_B}} \frac{k^4 \ dk}{(k^2+m^{* 2}_B)^{1/2}}
+ \frac{1}{3} \sum_l \frac{1}{\pi^2} 
\int_0^{K_{F_l}} \frac{k^4 \ dk}{(k^2+m^2_l)^{1/2}} 
+ \frac{\mu_{\nu_e}^4}{24\pi^2} ~.
\end{eqnarray}
Here, the last term is the contribution of neutrinos to the pressure.
The pressure due to antikaons is contained entirely in the meson
fields via their field equations (8)-(12).

At the interior of (proto)neutron stars, baryons are in chemical equilibrium 
under weak processes. Therefore the chemical potentials of baryons 
and leptons are governed by the equilibrium conditions
\begin{equation}
\mu_i = b_i \mu_n - q_i \left(\mu_e - \mu_{\nu_e}\right) ~,
\end{equation}
where $\mu_i$, $\mu_n$, $\mu_e$, and ${\mu_{\nu_e}}$ are respectively 
the chemical potentials of the ith baryon, neutrons, electrons, and neutrinos 
with 
$\mu_{i} = (k^2_{F_{i}} + m_i^{* 2} )^{1/2} + g_{\omega i} \omega_0
+ g_{\phi i} \phi_0 + I_{3i} g_{\rho i} \rho_{03}$ and $b_i$ and $q_i$ are
baryon and electric charge of ith baryon respectively. 
In neutron stars, electrons are converted to muons by 
$e^- \to \mu^- + \bar \nu_\mu + \nu_e$ when the electron chemical potential 
becomes equal to the muon mass. Therefore, we have $\mu_e=\mu_{\mu}$ in a
neutron star. On the other hand, muons are absent in a protoneutron star.
With the onset of $\bar K$ condensation, various strangeness changing 
processes may occur in (proto)neutron stars such as, 
$N \rightleftharpoons N + \bar K$ and $e^- \rightleftharpoons K^- + \nu_e$,
where $N\equiv (n,p)$ and $\bar K \equiv (K^-, \bar K^0)$ denote the 
isospin doublets for nucleons and antikaons, respectively. The
requirement of chemical equilibrium yields
\begin{eqnarray}
\mu_n - \mu_p &=& \mu_{K^-} = \mu_e - \mu_{\nu_e}~, \\
\mu_{\bar K^0} &=& 0 ~,
\end{eqnarray}
where $\mu_{K^-}$ and $\mu_{\bar K^0}$ are respectively the chemical
potentials of $K^-$ and $\bar K^0$. The above conditions determine the onset of 
antikaon condensations in the neutrino trapped matter. When the effective 
energy of $K^-$ meson ($\omega_{K^-}$) 
equals to its chemical potential ($\mu_{K^-}$) which, in
turn, is equal to $\mu_e-\mu_{\nu_e}$, a $K^-$ condensate 
is formed. Similarly, $\bar K^0$ condensation is formed when its in-medium 
energy satisfies the condition $\omega_{\bar K^0} = \mu_{\bar K^0} = 0$. 
It is to be noted here that the neutrino chemical potential ($\mu_{\nu_e}$) is
zero for neutrino free case corresponding to neutron stars.
For (proto)neutron star matter we need to include also the 
charge neutrality condition, which in the presence of antikaon condensate is
expressed as
\begin{equation}
\sum_i {q_i n_i} - n_{K^-} - n_e  - n_\mu = 0 ~,
\end{equation}
where $q_i$ and $n_i$ are the electric charge and density of ith baryon 
respectively. The other constraint in protoneutron stars is the number of 
leptons per baryon. Gravitational core collapse calculations of massive
stars indicate that the lepton fraction at the onset of trapping is
$Y_{L_e} = Y_e + Y_{\nu_e} \simeq 0.4$ and it is conserved on a dynamical time
scale \cite{Pra97}.

\section{Results and discussions}

In the effective field theoretic approach adopted here, we first consider 
nucleon-only matter where two distinct sets
of coupling constants for nucleons and kaons associated
with the exchange of $\sigma$, $\omega$, and $\rho$ mesons are required.
The nucleon-meson coupling constants generated by reproducing the
nuclear matter saturation properties are taken from Glendenning and 
Moszkowski of Ref. \cite{Gle91}. This set is referred to as GM1 and
listed in Table I.

Now we determine the kaon-meson coupling constants. According to
the quark model and isospin counting rule, the vector coupling constants are
given by
\begin{equation}
g_{\omega K} = \frac{1}{3} g_{\omega N} ~~~~~ {\rm and} ~~~~~
g_{\rho K} = g_{\rho N} ~.
\end{equation}
The scalar coupling constant is obtained from the real part of the 
$K^-$ optical potential at normal nuclear matter density 
\begin{equation}
U_{\bar K} \left(n_0\right) = - g_{\sigma K}\sigma - g_{\omega K}\omega_0 ~.
\end{equation}
The negative sign in the vector meson potential is due to G-parity.
The critical density of $\bar K$ condensation should therefore strongly
depend on the $K^-$ optical potential. 

It has been demonstrated in various calculations that antikaons feel an
attractive potential in normal nuclear matter \cite{Li,Fri94,Fri99,Koc,Waa}.
The analysis of $K^-$ atomic data \cite{Fri99} in a hybrid model
comprising of the relativistic mean field approach in the nuclear interior
and a phenomenological density dependent potential at low density, revealed
that the real part of the antikaon optical potential could be as large as
$U_{\bar K}=-180 \pm 20$ MeV at normal nuclear matter density and repulsive at 
low density in accordance with the low density theorem. In the coupled 
channel calculation \cite{Koc} for antikaons, the attractive antikaon 
potential depth was estimated to be $U_{\bar K}=-100$ MeV, whereas the chirally
motivated coupled channel approach predicted a depth of 
$U_{\bar K}=-120$ MeV \cite{Waa}. The wide range of values of the antikaon
potential depth as found in various calculations may be attributed to the 
different treatments of 
$\Lambda(1405)$-resonance which is considered to be an unstable
${\bar K} N$ bound state just below the $K^- p$ threshold. 
Therefore, we determine
the $K-\sigma$ coupling constant $g_{\sigma K}$ from a set of values of 
$U_{\bar K}(n_0)$ starting from $-100$ MeV to $-180$ MeV. This is listed
in Table II for the set GM1. Since the $\omega$-meson potential for $\bar K$
in this model is $V_\omega^K(n_0) = - g_{\omega K} \omega_0 \approx - 72$ MeV,
a rather large sigma-kaon coupling constant of $g_{\sigma K} = 3.674$ is
required to reproduce a depth of $-180$ MeV. It is to be noted that for this 
large depth, the value of the scalar coupling is similar to the prediction in 
the simple quark model i.e., $g_{\sigma K} = g_{\sigma N}/3$.
In an alternative approach the kaon-meson coupling constants were also
determined from the $s$-wave kaon-nucleon ($KN$) scattering length 
\cite{Sch,Coh,Bar}.

We now present results for (proto)neutron star matter containing 
nucleons, leptons and
$\bar K$ condensates for the parameter set GM1 in Tables I and II. 
In Figure 1, the scalar and vector potentials are displayed as a function 
of baryon density normalized to the equilibrium value of $n_0=0.153 fm^{-3}$ 
with the $K^-$ optical potential depth of $U_{\bar K}(n_0)=-160$
MeV for the neutrino free (top panel) and neutrino trapped (bottom panel) cases.
In both cases, the scalar ($\sigma$) and vector ($\omega$) potentials increase 
with density before the onset of antikaon condensation. 
In the neutrino free case, two 
curves touch each other just at the onset of $K^-$ condensation, whereas the
curve of the vector ($\omega$) potential crosses the curve of the scalar 
potential before
the appearance of antikaon condensation in the neutrino trapped case.  
After the formation of $\bar K$
condensates, the curves change slope i.e. the rate of increase of the fields
is altered with respect to the previous situations. With the appearance of 
$\bar K^0$ condensate, the isovector potential approaches zero with increasing
density in the neutrino free case, whereas it goes to zero and then bounces
back for the neutrino trapped case. It is found that the scalar potential for
the neutrino free matter becomes larger after the onset of antikaon 
condensation than that of the neutrino trapped matter. It follows from Eq. (8)
and the reasoning that the onsets of $\bar K$ condensation are delayed to 
higher densities in the neutrino trapped matter. 
Though the vector
potentials are comparable in both the cases up to $\sim 2.5 n_0$, it is higher
for the neutrino trapped case at higher densities. On the other hand, the 
isovector potential in the neutrino trapped matter is always smaller compared
with that of the neutrino free matter. Those variations in the meson fields may 
be attributed to the different behaviour of the source terms in the
field equations of motion (Eqs. (8),(10) and (12)) and the composition of
matter in two cases. 

The effective mass ratio of antikaons, $m^*_K/m_K$, are shown in Figure 2 as a 
function of normalised baryon density for $U_{\bar K}=-160$ MeV. 
The solid line represents the neutrino free case,
whereas the dotted line denotes the neutrino trapped situation. It is found
that the effective mass in the neutrino free matter is smaller 
than that of the neutrino trapped matter at higher densities. This may be 
attributed to the larger scalar potential at higher densities in the 
former case.

Figure 3 shows the s-wave antikaon condensation energies for the neutrino free
and neutrino trapped nuclear matter as a function of baryon density. 
The calculation is done with the antikaon potential depth of $U_{\bar K}=-160$
MeV. The solid 
lines correspond to the energy of $K^-$ meson, $\omega_{K^-}$, whereas the 
dashed lines indicate that of $\bar K^0$ meson, $\omega_{\bar K^0}$.
Also, the electron chemical potential ($\mu_e$) for 
the neutrino free case and the difference between electron and neutrino
chemical potentials, $\mu_e - \mu_{\nu_e}$, for the neutrino trapped case 
in the absence of $\bar K$ condensate are depicted in the figure. 
In the neutrino trapped matter, the in-medium energy of $K^{-}$ meson
is lower compared with that of the neutrino free matter. This difference in
the energies of $K^{-}$  meson stems from different behaviours of the meson 
fields in two cases as is evident from Fig. 1. The threshold densities of 
$K^{-}$ condensation in
the neutrino free and neutrino trapped matter are 2.43$n_0$ and 3.07$n_0$,
respectively. In the presence of trapped neutrinos, the onset of $K^{-}$ 
condensation is 
shifted to higher density because the difference of electron and neutrino 
chemical potentials ($\mu_e - \mu_{\nu_e}$) intersects the 
$\omega_{K^-}$ curve at higher density. On the other hand, the threshold 
condition for $\bar K^0$ condensation is always $\omega_{\bar K} = 0$. 
Unlike the situation with $K^-$ meson, we find that the in-medium energy of
$\bar K^0$ meson in the neutrino trapped matter is higher than that of the
neutrino free case. The threshold
densities for $\bar K^0$ condensation are 3.59$n_0$ and 3.81$n_0$ for the 
neutrino free and neutrino trapped matter, respectively. The early appearance
of $K^-$ condensate delays the formation of $\bar K^0$ condensate in the 
neutrino free(trapped) matter to a higher density. Due to the presence
of neutrinos, the shift in the threshold density of $\bar K^0$ 
condensation is smaller with respect to the neutrino free case 
than the corresponding situation with $K^{-}$ condensation. 
We note that the difference between the energies of $\bar K^0$ and
$K^{-}$ in the neutrino trapped case is smaller than that of the neutrino
free case because the isovector potential is smaller in the former case.
Threshold densities of $\bar K$ condensation for the GM1 set and other values
of $U_{\bar K}(n_0)$ are given in Table III. The values given in the parentheses
correspond to the neutrino free matter.

The populations of various particles in proto(neutron) star matter with 
$K^-$ and $\bar K^0$ condensation for $U_{\bar K}=-160$ MeV
are shown in Figure 4. In the top panel we exhibit the particle abundances of 
the neutrino free star matter. In the neutrino 
free matter, once $K^-$ condensate sets in at $2.43n_0$,
it rapidly increases with density replacing the leptons in maintaining
the charge neutrality. The proton density  becomes equal to $K^-$ condensate 
density because of the charge neutrality.
With the onset of $\bar K^0$ condensate at 3.59$n_0$, the
neutron and proton abundances become identical resulting in a symmetric matter 
of nucleons and antikaons \cite{Pal}. In the previous calculation of $K^-$
condensation in neutron star matter \cite{Pra97,Ell,Kno}, it was found that
protons were more abundant than neutrons at higher densities. 
In the bottom panel of Fig. 4, we
show the particle fractions with both $K^-$ and $\bar K^0$ condensates for
the neutrino trapped matter. We have a somewhat different picture here. With
the formation of $K^{-}$ condensate at 3.07$n_0$, it can not replace electrons
totally like the neutrino free case because of the constraint $Y_{L_e}=0.4$ in 
the system. At higher baryon densities, the electron density slowly falls and 
the density of $K^{-}$ condensate becomes higher than the electron density. 
As soon as $\bar K^0$ condensate is formed at 3.81$n_0$, the neutron 
density becomes equal to the proton density 
and it continues in the high density regime. Like the
neutrino free case (top panel), the neutrino trapped matter becomes symmetric
nuclear matter just with the formation of $\bar K^0$ condensate. On the
other hand, the density of $\bar K^0$ condensate increases with baryon density 
uninterruptedly and even becomes larger than the density of $K^{-}$ condensate
beyond $\sim 4 n_0$. As a result, the $\rho$ meson field becomes zero at the
onset of $\bar K^0$ condensation and then bounces back (see Fig. 1) 
unlike the situation in the neutrino free matter where the isovector 
field is zero at and beyond the onset of $\bar K^0$ condensation. It is
interesting to note that the beta equilibrated and charge neutral neutrino 
trapped matter is mainly dominated by $\bar K^0$ condensate than $K^{-}$
condensate after the formation of $\bar K^0$ condensate.

The equation of state (EOS) or the pressure ($P$) versus energy density 
($\epsilon$) for the neutrino free and neutrino trapped matter is displayed
in Figure 5. The top panel represents the nucleon-only matter. Here, the 
dashed line stands for the neutrino trapped case and the solid line implies the
neutrino free case. The overall EOS of the neutrino trapped matter is softer
compared with that of the neutrino free matter. It is the delicate interplay 
of the contributions of the symmetry and lepton terms to the energy density
(pressure) in both cases. As trapped neutrinos leave the system, the conversion
of protons to neutrons increases the energy density (pressure) more than it is
decreased by the loss of neutrinos. This scenario is changed in the presence of
antikaon condensation. This is demonstrated in the bottom panel of
Fig. 5 with $U_{\bar K}=-160$ MeV. We find that the overall EOS 
of the neutrino trapped matter is now stiffer than that of the neutrino free
matter. During deleptonization, the contribution of the nuclear symmetry term
to the energy density (pressure) does not change at all in this case with
$K^-$ and $\bar K^0$ condensate at higher densities. 
This happens because both the neutrino free and neutrino
trapped matter become isospin saturated nuclear matter at the onset of 
$\bar K^0$ condensation. Such a situation does not arise for 
(proto)neutron star matter with $K^-$ condensate
because more protons than neutrons are favoured by $K^{-}$ condensate due to
the charge neutrality. Therefore, it is the lepton contribution which makes the
difference between the EOS of the neutrino free and neutrino trapped matter
including both $K^{-}$ and $\bar K^0$ condensate. It is important to note
here that the incompressibility of matter ($K= 9dP/dn_B$) becomes
negative with the formation of $\bar K$ condensates in the neutrino free matter
as is evident from Fig. 5. To get rid of those unphysical regions from the
energy density versus pressure curve and maintain a positive 
incompressibility, the Maxwell construction is done here. 
This implies a first order phase transition. We note that 
the phase transition is second order for lower values of the antikaon
potential. It is worth mentioning here that the condensation of $K^-$ meson
was treated as first order phase transition using Gibbs criteria \cite{Gle99}.
In this case, the matter would have a normal phase of baryons and leptons 
at low density followed by a mixed phase of $K^-$ condensate and baryons and 
a pure phase of $K^-$ condensate at higher densities. This scenario would be 
more complicated with further appearance of $\bar K^0$ condensate
because $K^-$ and $\bar K^0$ condensation have to be treated as two 
separate first order phase transitions. Therefore, the treatment of 
such a problem is beyond the scope of this work. 

Now we present the results of static structures of (proto)neutron stars 
calculated using Tolmann-Oppenheimer-Volkov (TOV) equations. The static 
(proto)neutron star sequences representing the stellar masses $M/M_\odot$ and
the corresponding central energy density $\varepsilon_c$ are shown in 
Figure 6 for the GM1 set. The neutrino free stars
are denoted by the solid lines and the neutrino trapped stars are represented
by the dashed lines in Fig. 6. 
The "conventional" scenario i.e. stars made of nucleons, leptons and
no $\bar K$ condensation is shown in the top panel. 
The maximum masses ($M_{max}$) and central densities ($n_c$) of the 
neutrino trapped and neutrino free stars are respectively
given by 2.283(2.364)$M_{\odot}$ and 5.84(5.63)$n_0$. The protoneutron star
has a smaller mass compared with that of the neutron star because the EOS is 
softer in the former case. 
In the bottom panel, the masses of (proto)neutron stars
with both $K^{-}$ and $\bar K^0$ condensation and $U_{\bar K}(n_0)=-160$ MeV
are plotted with central energy density. 
In this case, the maximum masses of the (proto)neutron stars are
1.97(1.55)$M_{\odot}$ corresponding to the central densities 4.89(3.59)$n_0$,
respectively. In this case, the maximum masses of the (proto)neutron
stars are smaller than those of the "conventional" scenario. This can be 
attributed to the softening of the EOS due to the presence of $\bar K$ 
condensates. In the previous calculations including $K^-$ condensate
\cite{Pra97,Bro92,Tho,Kno,Sch,Gle99} this kind of softening 
in the EOS and reduction in the maximum mass of the star was observed. 
The net result of $K^{-}$ condensation is that the maximum mass
stars contain more protons than neutrons at higher densities. 
Brown and collaborators \cite{Bro2,Brown} called those stars as 
"nuclear matter" stars rather than neutron stars. In the neutrino
trapped case, the additional softening in the EOS due to $\bar K^0$ condensate
results in a smaller maximum mass for the star compared with the corresponding 
situation with only $K^{-}$ condensate. 
In the bottom panel and Table III, we find that the threshold density of
$\bar K^0$ condensation for the neutrino trapped case lies well inside 
the central density of the maximum mass star. 
This implies that a significant region of the protoneutron star may
contain $\bar K^0$ condensate along with $K^{-}$ condensate. 
In the neutrino free case, the threshold density of $\bar K^0$ condensation 
coincides with the central density of the maximum mass star. 
It is interesting to note that the maximum mass stars in the presence of 
$K^{-}$ and $\bar K^0$ condensate contain exactly equal number
of protons and neutrons at higher densities. Therefore, those stars may be 
called as "symmetric nuclear matter" stars. We have also calculated maximum
masses and central densities of the (proto)neutron stars for the GM1 set and 
other values of $U_{\bar K}(n_0)$ and those are tabulated in Table III. 
The values given in the parentheses correspond to the neutrino free matter.

In a recent calculation for the neutrino free matter including $\bar K$ 
condensates,
it has been noted that the critical densities of antikaon condensation are
sensitive to the nuclear equation of state apart from its dependence on 
the antikaon potential depth \cite{Pal}. Therefore, we also study the formation
of $\bar K$ condensates in the neutrino trapped matter using a softer EOS.
In this calculation, the model Lagrangian density contains a nonlinear
$\omega$ meson term \cite{Bod,Sug} besides the self interaction term for 
the scalar meson. The form of the nonlinear $\omega$ meson term is, 
\begin{equation}
{\cal L}_{\omega^4} = \frac{1}{4} g_4 \left(\omega_\mu \omega^\mu \right)^2 ~.
\end{equation}
Sugahara and Toki \cite{Sug} showed that such a model agreed with the
relativistic Brueckner Hartree Fock results reasonably well. The parameters
of the model were obtained by fitting the experimental data for binding
energies and charge radii of heavy nuclei. This set of parameters is known as
the TM1 set \cite{Sug}. The nucleon-meson couplings of the TM1 set are shown in 
Table I and $g_{\sigma K}$ couplings for various $U_{\bar K}(n_0)$ are given
in Table II. Now we present the results for the calculation of 
$\bar K$ condensation in the neutrino
trapped nucleon-only matter using the TM1 set. The critical densities of 
$\bar K$ condensation, maximum masses of (proto)neutron stars 
with their corresponding central 
densities for the TM1 set are shown in Table IV. The values in the parentheses
correspond to the neutrino free cases. Because of the presence
of the nonlinear $\omega$ meson term, the TM1 set results in a softer EOS
than that with the GM1 set. As a result, the critical densities of $\bar K$
condensation for the TM1 set are shifted to higher densities compared with those
of the GM1 set. It is found that $\bar K^0$ condensation is formed inside the
maximum mass neutron stars for $U_{\bar K}(n_0)\geq -160$ MeV and inside the
maximum mass protoneutron star only for $U_{\bar K}(n_0)=-180$ MeV in the 
TM1 set. It is important to note that the phase transition is of second order
for all values of the antikaon potential depth in the TM1 set \cite{Pal}.

Now we discuss the situations when hyperons are included in the calculation
in addition to nucleons. It was noted earlier that the presence of hyperons 
delayed the onsets of $\bar K$ condensation to much higher density 
\cite{Pra97,Ell,Mut,Kno,Sch,Pal}. Since the core of a (proto)neutron star
may be hyperon rich, we include hyperon-hyperon interaction besides 
hyperon-nucleon interaction in our calculation. This is accounted by 
considering two additional hidden-strangeness mesons - the scalar meson
$f_0$(975) (denoted by $\sigma^*$) and the vector meson $\phi$(1020).
The vector coupling constants for hyperons are determined from the SU(6) 
symmetry as,
\begin{eqnarray}
\frac{1}{2}g_{\omega \Lambda} = \frac{1}{2}g_{\omega \Sigma} = g_{\omega \Xi} = 
\frac{1}{3} g_{\omega N},\nonumber\\
\frac{1}{2}g_{\rho \Sigma} = g_{\rho \Xi} = g_{\rho N} ~~~ {\rm ;}~~~ 
g_{\rho \Lambda} = 0, \nonumber\\
2 g_{\phi \Lambda} = 2 g_{\phi \Sigma} = g_{\phi \Xi} = 
-\frac{2\sqrt{2}}{3} g_{\omega N}. ~ 
\end{eqnarray}
The scalar meson ($\sigma$) coupling to hyperons is obtained from the 
potential depth of a hyperon (Y) in the saturated nuclear matter
\begin{equation}
U_Y^N(n_0) = - g_{\sigma Y} {\sigma} + g_{\omega Y} {\omega_0}.
\end{equation}
The analysis of energy levels in $\Lambda$-hypernuclei suggests a well depth of 
$\Lambda$ in symmetric matter $U_{\Lambda}^N(n_0)=-30$ MeV \cite{Chr,Dov}. 
On the other hand, recent analysis of 
a few $\Xi$-hypernuclei events predicts a $\Xi$ well depth of
$U_{\Xi}^N(n_0)=-18$ MeV \cite{Fak,Kha}. 
However, the situation for the $\Sigma$ potential in 
normal matter is very unclear. The only known bound $\Sigma$-hypernuclei is the
light system $^4_{\Sigma}He$ \cite{Hay}. The most updated analysis 
of $\Sigma^-$ atomic data indicates a strong isoscalar repulsion 
in the $\Sigma$-nuclear matter interaction \cite{Fri94}.
Therefore, we use a repulsive $\Sigma$ well depth of 
$U_{\Sigma}^N(n_0)=30$ MeV \cite{Fri94} in our calculation.

The $\sigma^*$-Y coupling constants are obtained by fitting them to a well
depth , ${U_{Y}^{(Y^{'})}}{(n_0)}$, for a hyperon (Y) in a hyperon ($Y^{'}$) 
matter at nuclear saturation density \cite{Sch,Sch94}. It is given as
\begin{equation}
U_{\Xi}^{(\Xi)}(n_0) = U_{\Lambda}^{(\Xi)}(n_0) = 2 U_{\Xi}^{(\Lambda)}(n_0) 
= 2 U_{\Lambda}^{(\Lambda)}(n_0) = -40~MeV.
\end{equation} 
It is to be noted that nucleons do not couple to the strange mesons i.e.
$g_{{\sigma^*} N}=g_{\phi N}=0$.

The strange meson fields also couple with (anti)kaons. Following 
Ref.\cite{Sch}, the $\sigma^*$-K coupling constant is determined from the
decay of $f_0$(925) as $g_{\sigma^*K}=2.65$, whereas the vector $\phi$ meson
coupling with (anti)kaons is obtained from the SU(3) relation as 
$\sqrt{2} g_{\phi K} = 6.04$. 

Switching off Y-Y interactions and antikaon condensation, we study the 
neutrino free(trapped) hyperon matter for the GM1 and TM1 set and for the
above choices of hyperon-nucleon coupling constants. Here, we find that 
$\Lambda$ hyperon is the first strange baryon to appear in the neutrino
free(trapped) matter. It is closely followed by $\Xi^-$ hyperons. Because of 
the repulsive $\Sigma$-nucleon interaction, $\Sigma$ hyperons do not appear in 
the systems. With the appearance of hyperons , equations of state in the 
neutrino free(trapped) matter are softer compared with those of the  
nucleon-only matter excluding $\bar K$ condensation.
The maximum masses and their corresponding central densities are given
in the last rows of Table III and IV. The values in the parentheses 
correspond to the neutrino free cases. Like the situations with antikaon 
condensates in the neutrino free(trapped) nucleon-only
matter, the maximum mass in the neutrino trapped hyperon matter is larger than 
that of the corresponding neutrino free case as found in the previous 
calculations \cite{Pra97,Ell,Kno}.   

Now we study the formation of antikaon condensation in neutrino free(trapped)
matter including hyperon-hyperon interaction for the GM1 set. Earlier we 
have found that a softer EOS shifts the threshold densities of antikaon
condensations to higher densities. Also, the presence of hyperons which make
the EOS softer, delays the onsets of $\bar K$ condensation to higher densities 
\cite{Pra97,Ell,Kno,Sch,Pal}. In this case, we find that antikaon 
condensations do not occur even at densities as large as 7.5$n_0$ 
in the neutrino free(trapped) hyperon
matter for $U_{\bar K}(n_0) < -160$ MeV. Antikaon condensations appear in
hyperon rich matter with and without neutrinos for $U_{\bar K}(n_0)\geq -160$
MeV. The particle abundances in the presence of hyperons and antikaon 
condensations are shown in Figure 7 for $U_{\bar K}=-160$ MeV. The top panel 
depicts the neutrino free case whereas the neutrino trapped case is shown 
in the bottom panel. Here, we note that $K^-$ and $\bar K^0$ condensation set 
in at 2.48(3.34)$n_0$ and 4.06(4.18)$n_0$ for the neutrino free(trapped) matter,
respectively. Also, we find that only $\Lambda$ hyperons appear both in the 
neutrino free and trapped matter. The $\Sigma$s are excluded from the systems
because of the repulsive $\Sigma$-nucleon interaction. On the other hand, the
early appearance of $K^-$ condensation  in the neutrino free(trapped) matter
suppresses the appearance of $\Xi^-$ hyperons because it is energetically
favourable for $K^-$ condensate to maintain charge neutrality in the
systems. It is worth mentioning here that $\bar K$ condensates play more
dominant role than hyperons in determining various properties of
the (proto)neutron stars in this case.

In Figure 8, we display the EOS (P vs. $\epsilon$) 
in the top panel and the mass sequence of the (proto)neutron stars in the 
bottom panel for the GM1 set and $U_{\bar K}(n_0)=-160$ MeV. 
The neutrino trapped matter
has a stiffer EOS than that of the corresponding neutrino free matter.
The maximum masses and their central densities for the neutrino free(trapped)
cases are 1.57(1.98)$M_{\odot}$ and 4.48(5.24)$n_0$ respectively.

The delayed neutrino emission and the possible 
black hole formation in the context of
SN 1987A had been extensively discussed by Bethe and Brown \cite{Bro2,Brown}.
Assuming SN 1987A has collapsed into a black hole, Bethe and Brown estimated the
gravitational mass for the compact remnant in it to be 1.56$M_{\odot}$ from
the Ni production. This maximum mass of the cold compact object is known as
the Bethe-Brown limit \cite{Brown}. They argued that progenitor stars 
having masses in the range 18-30 $M_{\odot}$ would first explode as 
supernovae and later the compact objects go into low mass black holes
returning matter to the galaxy \cite{Bro2,Brown}.

In the previous studies \cite{Bro2,Brown,Pra97}, it was shown that the
protoneutron stars become unstable after deleptonization and cooling. 
The metastability occurs when exotic matter such as hyperons or 
antikaon condensates appears during the evolution of the protoneutron stars.
Bethe-Brown exploited the idea of the formation of  $K^{-}$ condensation 
in dense nuclear matter \cite{Brown} to understand the metastability of 
the compact object during the evolution of the protoneutron 
stars to low mass black holes. In the "conventional" scenario without antikaon
condensation or hyperons, the "window" i.e. the difference of the maximum 
masses in the
neutrino free and neutrino trapped matter consisting of nucleons is small. 
Later it was shown 
that the inclusion of thermal pressure could raise the maximum mass
of the protoneutron star slightly above that of the neutron star \cite{Pra97}. 
However, there is a different scenario with $K^{-}$ condensation as noted by 
various authors \cite{Bro2,Brown,Pra97,Tho}. They found
that mostly the lepton pressure could stabilize larger mass in the neutrino
trapped case during the evolution. 
In our calculation with nucleons, leptons and $K^{-}$ condensation, 
the "window" is $\sim 0.5 M_{\odot}$ for the 
GM1 set and the antikaon potential depth of -160 MeV. With further inclusion 
of $\bar K^0$ condensate, the high density matter 
contains exactly as many protons as neutrons. As a result, the high density
matter remains the symmetric nuclear matter even after the deleptonization and 
cooling. In the presence of both the condensates, it is exactly the lepton 
pressure which stabilizes larger maximum mass of the protoneutron star for 
short times. In this case, we obtain 
the maximum mass window of $\sim 0.4 M_{\odot}$. 
In our model calculation with the GM1 set and the antikaon potential 
of $U_{\bar K}(n_0)=-160$ MeV, the protoneutron stars consisting of nucleons, 
leptons and $K^-$ and $\bar K^0$ condensate have a maximum mass 
$\sim 2 M_{\odot}$ which could be stable during the deleptonization and cooling.
As the trapped neutrinos leave the system, the lepton pressure in the core 
decreases. At the same time, the core of the nascent star is heated up and 
attains a higher value of entropy \cite{Bur}. Recently, it has been shown by 
Pons et al. \cite{Pon} that this thermal effect on the maximum mass is 
comparable to that of the trapped neutrinos. As a consequence, the compact 
object would be stable for much longer duration.
The inclusion of the thermal pressure in the calculation would again stabilize 
an additional mass of $\sim 0.1 M_{\odot}$. 
As the system cools down, the compact object with a maximum mass 
$\sim 2 M_{\odot}$ which would be stable for short times in our calculation,
is larger than that of the stable cold mass and the Bethe-Brown limit (1.56 
$M_{\odot}$). Consequently, it would collapse into a low mass black hole. 
We retain the same qualitative feature of the metastability of the protoneutron 
stars including hyperons along with antikaon condensation.
       
\section{Summary and Conclusions}

We have studied antikaon condensation putting emphasis on the formation of
$\bar K^0$ condensation in the neutrino trapped nuclear and hyperon matter
within the relativistic mean field models. The baryon-baryon and 
(anti)kaon-baryon
interactions are treated on the same footing in this work. Those interactions
are mediated by the exchange of mesons. Two different model Lagrangians are
adopted in this calculation. The model Lagrangian which contains the scalar
self-interaction term, is characterised by the GM1 parameter set. Besides the
scalar self-interaction term, the other model Lagrangian includes the 
non-linear $\omega$ meson term and the corresponding parameter set is denoted
by the TM1 set. It is found that the threshold densities of 
antikaon condensations are sensitive to the
equation of state and the antikaon potential in normal nuclear matter. The
calculations performed with the GM1 and TM1 set show that $K^-$ condensation
always happens earlier than $\bar K^0$ condensation. The threshold densities
of $\bar K$ condensation in the TM1 set are higher than those of the GM1 set
because the equation of state is softer in the former case. It is found that
the threshold densities of antikaon condensation in the neutrino trapped
nucleon-only matter are higher than those of the corresponding neutrino free
case. In the presence of neutrinos, the shift in the threshold density of
$K^-$ condensation with respect to the neutrino free case is higher 
than that of $\bar K^0$ condensation. 

With the onset of $\bar K^0$ condensation, abundances of neutrons and protons
become equal and the density of $\bar K^0$ condensate 
increases rapidly and becomes larger than that of $K^-$ condensate in the 
neutrino trapped nuclear matter. On the contrary, the neutrino free nuclear
matter is not only symmetric in neutrons and protons but also in $K^-$ and 
$\bar K^0$ mesons at high density. Therefore, it is possible that $\bar K^0$
condensate in the neutrino trapped matter would play a dominant role over
$K^-$ condensate at higher densities. Unlike the situation in the neutrino
free nuclear matter, $K^-$ mesons in the condensate replace electrons
partially in the neutrino trapped matter because of the lepton number 
constraint in the system. The presence of $K^-$ and 
$\bar K^0$ condensate in the neutrino trapped nuclear matter makes the overall
equation of state softer compared with the situation without antikaon 
condensate. We find that $\bar K^0$ condensate is formed well inside the 
maximum mass protoneutron stars for higher values of the antikaon potential in 
the calculation using the TM1 set. On the other hand, the calculation with the
GM1 set implies that $\bar K^0$ condensate occupies a significant region of the
maximum mass stars for rather smaller values of the antikaon potential. In the
presence of hyperons, antikaon condensate is formed for large values of
$U_{\bar K} \ge -160$ MeV in the GM1 set. In this case, it is found that only 
$\Lambda$ hyperons appear in the neutrino free(trapped) hyperon matter. 
Therefore $K^-$ and $\bar K^0$ condensate may play the most dominant 
role in determining various gross properties of (proto)neutron stars 
than hyperons. 

We have revisited the scenario of the metastability of protoneutron stars and
their evolution to low mass black holes in the context of the calculation
of $\bar K^0$ condensation along with $K^-$ condensation in the neutrino
free(trapped) nuclear and hyperon matter. It is found that the maximum mass
of a protoneutron star is larger than that of the corresponding neutron star
and also the Bethe-Brown limit of $1.56 M_{\odot}$
for a neutron star. Therefore, the protoneutron star would be stable during the
deleptonization and cooling and later it may collapse into a low mass black 
hole.

\vspace{0.5cm}
\leftline{Acknowledgment}

We acknowledge many fruitful discussions with Subrata Pal.


\begin{table}

\caption{The nucleon-meson coupling constants in 
the GM1 set are taken from Ref. [22]. In this relativistic model, the baryons
interact via nonlinear $\sigma$-meson and linear $\omega$-meson exchanges. 
The coupling constants are obtained by reproducing the nuclear matter 
properties of binding energy $E/B=-16.3$ MeV, baryon density $n_0=0.153$ 
fm$^{-3}$, asymmetry energy coefficient $a_{\rm asy}=32.5$ MeV, 
incompressibility $K=300$ MeV, and effective nucleon mass $m^*_N/m_N = 0.70$. 
The hadronic masses are $m_N=938$ MeV, $m_\sigma=550$ MeV, 
$m_\omega=783$ MeV, and $m_\rho=770$ MeV. The parameter set
TM1 is obtained from Ref. [30] which incorporates nonlinear exchanges in 
both $\sigma$ and $\omega$ mesons. The nuclear matter properties in the TM1 set
are $E/B=-16.3$ MeV, $n_0=0.145$ fm$^{-3}$, $a_{\rm asy}=36.9$ MeV, 
$K=281$ MeV, and $m^*_N/m_N = 0.634$. All the hadronic masses in 
this model are same as GM1 except for $\sigma$-meson which is 
$m_\sigma=511.198$ MeV. All the parameters are dimensionless, except $g_2$ 
which is in fm$^{-1}$.}

\begin{tabular}{ccccccc} 

\hfil& $g_{\sigma N}$& $g_{\omega N}$& $g_{\rho N}$&
$g_2$& $g_3$& $g_4$ \\ \hline
GM1& 9.5708& 10.5964& 8.1957& 12.2817& -8.9780& $-$ \\
TM1& 10.0289& 12.6139& 4.6322& -7.2325& 0.6183& 71.3075 \\

\end{tabular}
\end{table}

\begin{table}

\caption{The coupling constants for antikaons ($\bar K$) to
$\sigma$-meson, $g_{\sigma K}$, for various values of $\bar K$ optical 
potential depths $U_{\bar K}$ (in MeV) at the saturation density. The 
results are for the GM1 and TM1 set.}

\begin{tabular}{cccccc} 

$U_{\bar K}$& -100& -120& -140& -160& -180 \\ \hline
GM1& 0.9542& 1.6337& 2.3142& 2.9937& 3.6742\\
TM1& 0.2537& 0.8384& 1.4241& 2.0098& 2.5955 \\

\end{tabular}
\end{table}

\begin{table}

\caption{The maximum masses $M_{max}$ and their corresponding central 
densities $u_{cent}$
=$n_{cent}$/$n_{0}$ for the neutrino-trapped nucleon-only (np) star matter and 
for stars with 
further inclusion of hyperons (npH) are given below. The lepton fraction
in the neutrino trapped matter is $Y_{L_e} = Y_e + Y_{\nu_e} = 0.4$. 
The results are for the GM1 set. 
For protoneutron star matter with nucleons and antikaons (np$\bar{K}$), the 
critical densities for $K^{-}$ and $\bar{K}^0$ condensation, 
$u_{cr}(K^{-})$ and $u_{cr}(\bar{K}^0)$, and also the results for 
$M_{max}$ and $u_{cent}$ at various values of the antikaon potential depth 
$U_{\bar{K}}$ (in MeV) at the 
saturation density are given. The values in the parentheses are for 
the neutrino free matter relevant to neutron stars.}

\begin{tabular}{cccccc}

{}&${U_{\bar{K}} }$&${u_{cr}(K^{-})}$&${u_{cr}(\bar{K}^{0})}$&${u_{cent}}$
&${M_{max}/M_{\odot}}$\\ \hline
{np}&{-}&{-}&{-}&{5.84~(5.63)}&{2.283~(2.364)}\\
&&&&&\\
{}&{-100}&{4.40~(3.45)}&{5.71~(5.51)}&{5.68~(5.17)}&{2.258~(2.211)}\\
{}&{-120}&{3.90~(3.05)}&{5.03~(4.83)}&{5.57~(5.19)}&{2.218~(2.077)}\\
${np\bar{K}}$&{-140}&{3.45~(2.71)}&{4.39~(4.19)}&{5.33~(4.75)}&{2.134~(1.856)}\\
{}&{-160}&{3.07~(2.43)}&{3.81~(3.59)}&{4.89~(3.59)}&{1.970~(1.551)}\\
{}&{-180}&{2.74~(2.19)}&{3.31~(3.07)}&{4.74~(3.09)}&{1.686~(1.217)}\\
&&&&&\\
{npH}&{-}&{-}&{-}&{5.66~(5.16)}&{2.043~(1.789)}\\ 
\end{tabular}
\end{table}

\begin{table}

\caption{Same as Table III, but for the TM1 set.}

\begin{tabular}{cccccc}
{}&${U_{\bar{K}} }$&${u_{cr}(K^{-})}$&${u_{cr}(\bar{K}^{0})}$&${u_{cent}}$
&${M_{max}/M_{\odot}}$\\ \hline
{np}&{-}&{-}&{-}&{6.14~(5.97)}&{2.099~(2.179)}\\
&&&&&\\
{}&{-100}&{6.80~(4.15)}&{11.61~(11.12)}&{6.14~(5.67)}&{2.099~(2.142)}\\
{}&{-120}&{5.63~(3.55)}&{9.38~(9.13)}&{6.13~(5.55)}&{2.098~(2.083)}\\
${np\bar{K}}$&{-140}&{4.68~(3.05)}&{7.53~(7.43)}&{6.02~(5.65)}&{2.087~(1.986)}\\
{}&{-160}&{3.92~(2.67)}&{6.04~(5.99)}&{5.92~(6.37)}&{2.058~(1.857)}\\
{}&{-180}&{3.34~(2.37)}&{4.85~(4.81)}&{5.72~(6.01)}&{1.985~(1.641)}\\
&&&&&\\
{npH}&{-}&{-}&{-}&{5.75~(4.88)}&{1.918~(1.733)}\\
\end{tabular}
\end{table}
 \newpage 
\vspace{-2cm}

{\centerline{
\epsfxsize=14cm
\epsfysize=17cm
\epsffile{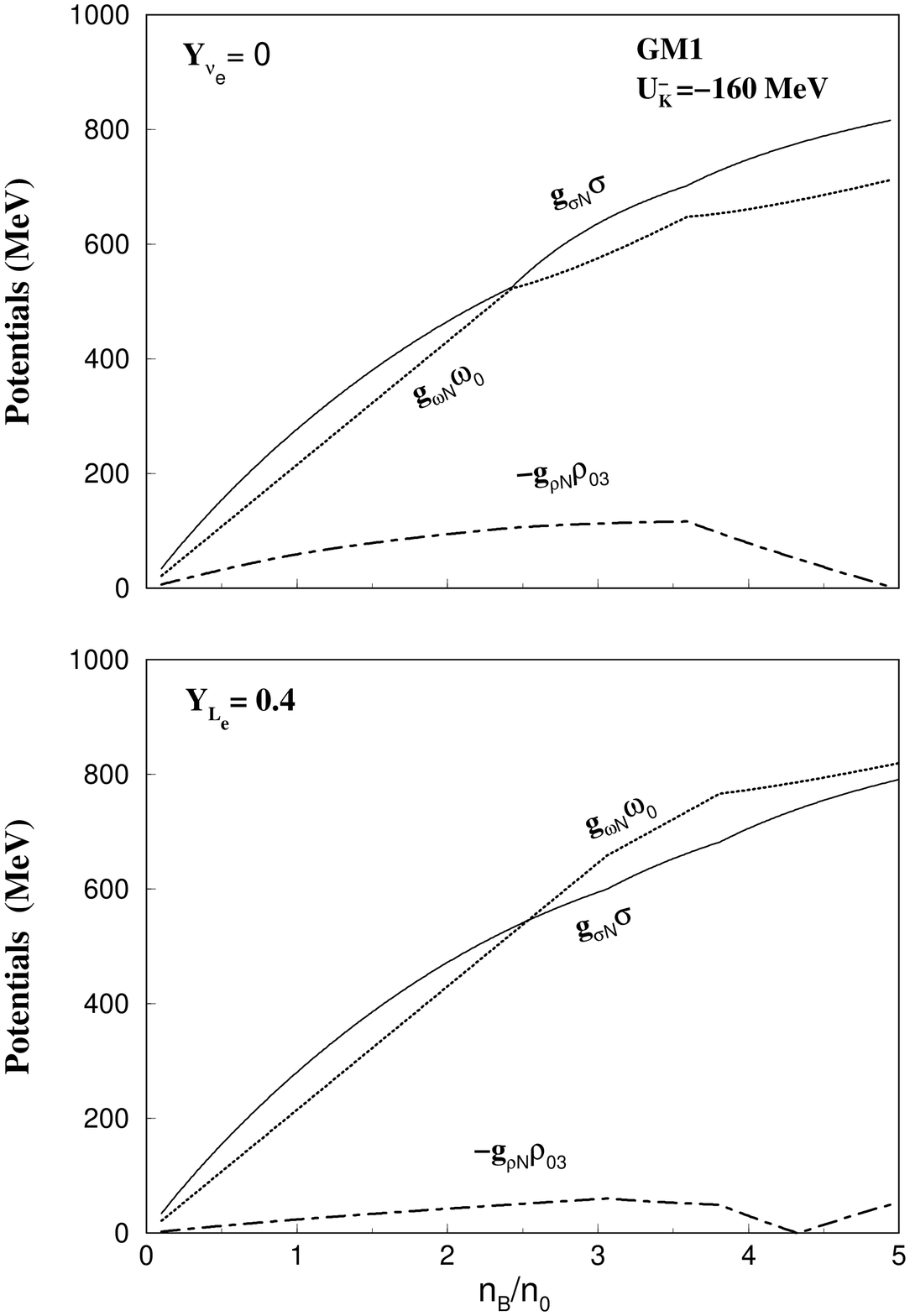}
}}

\vspace{1cm}

\noindent{\small{
FIG. 1. The mean meson potentials 
versus the baryon density $n_B/n_0$ in the GM1 set for the 
neutrino free (top panel) and neutrino trapped (bottom panel) nucleon-only star 
matter with the inclusion of antikaon, $K^-$ and $\bar K^0$, condensation.
The $\bar K$ optical potential depth is $U_{\bar K}=-160$ MeV at the normal 
nuclear matter density of $n_0=0.153$ fm$^{-3}$.}} 
 \newpage
\vspace{-2cm}

{\centerline{
\epsfxsize=14cm
\epsfysize=17cm
\epsffile{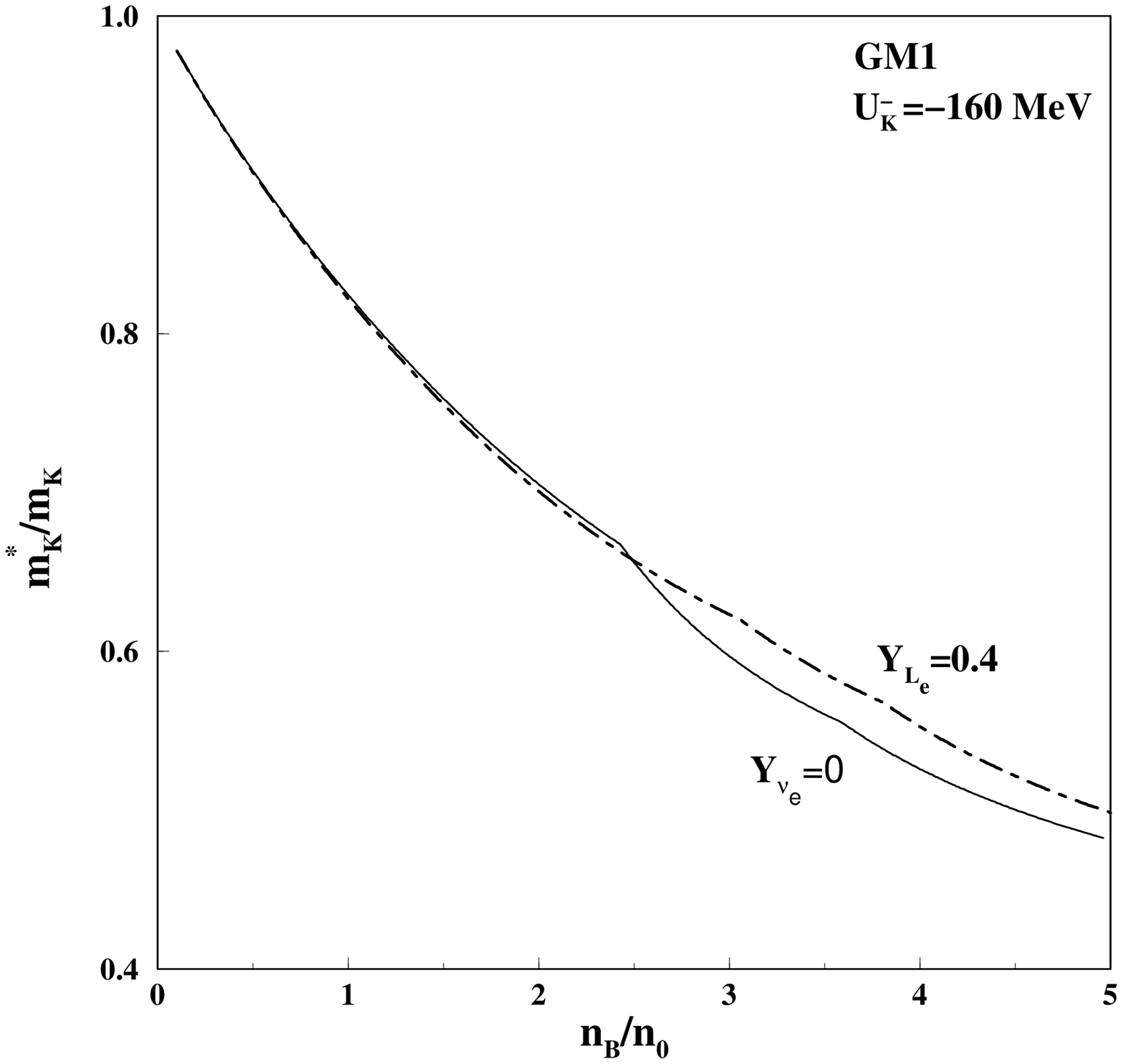}
}}

\vspace{1cm}

\noindent{\small{
FIG. 2. The variation of the effective mass of (anti)kaons $m_K^*/m_K$ as a 
function of baryon density $n_B/n_0$ for the neutron star matter (solid line) 
and protoneutron star matter (dashed line) consisting of nucleons and antikaon
condensates in the GM1 set. The $\bar K$ optical potential depth at normal 
nuclear matter density is  $U_{\bar K}= -160$ MeV for this calculation.}}
 \newpage
\vspace{-2cm}

{\centerline{
\epsfxsize=14cm
\epsfysize=17cm
\epsffile{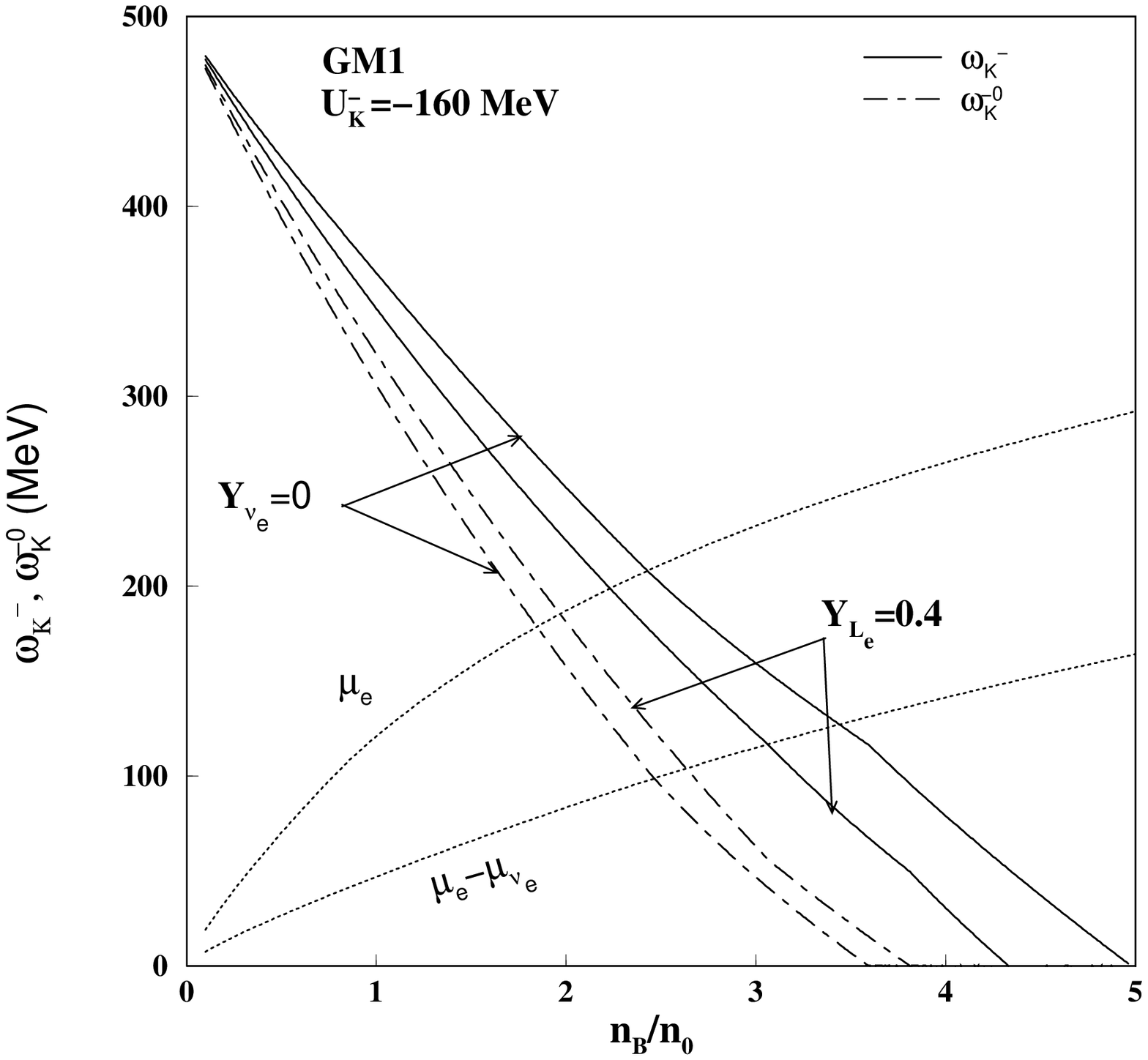}
}}

\vspace{1cm}

\noindent{\small{
FIG. 3. The in-medium energy of $K^-$ (solid lines) and $\bar K^0$ 
(dashed lines) versus baryon density for the neutrino free(trapped) nucleon-only
matter in the GM1 set. The electro chemical potential ($\mu_e$) and the
difference of the electro chemical potential and the neutrino chemical 
potential i.e. $\mu_e - \mu_{\nu_e}$ are also shown in the figure. 
The $\bar K$ optical potential at normal nuclear matter density is 
$U_{\bar K} = -160$ MeV.}}

 \newpage
\vspace{1cm}

{\centerline{
\epsfxsize=14cm
\epsfysize=17cm
\epsffile{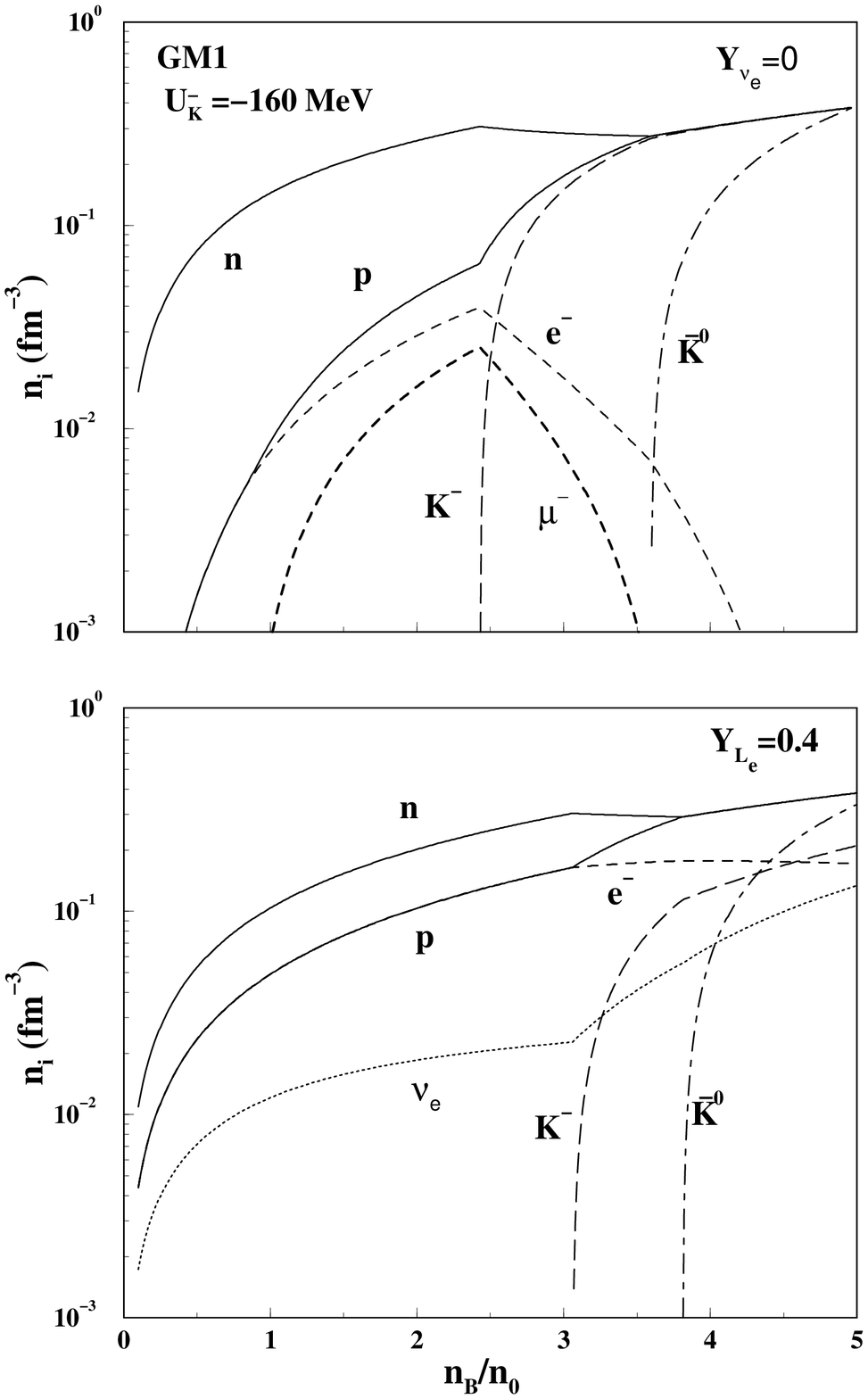}
}}

\vspace{1cm}

\noindent{\small{
FIG. 4. The proper number densities $n_i$ of various compositions in 
the (proto)neutron star matter including antikaon condensates in the GM1 model. 
The results are for the neutrino free matter (top panel) and neutrino 
trapped  matter (bottom panel). The $\bar K$ optical potential 
at normal nuclear matter density is $U_{\bar K} = -160$ MeV.}}

 \newpage
\vspace{-2cm}

{\centerline{
\epsfxsize=14cm
\epsfysize=17cm
\epsffile{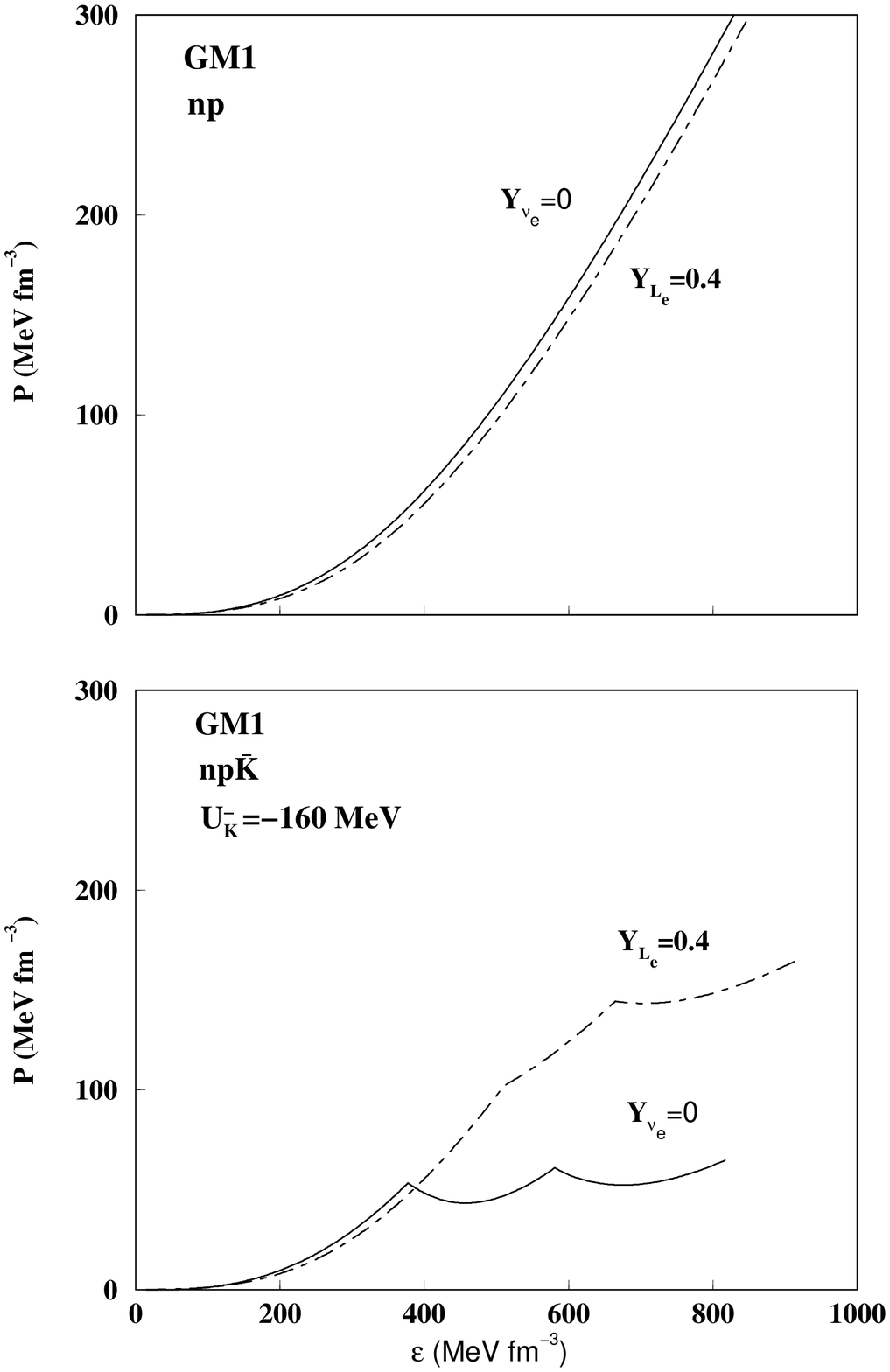}
}}

\vspace{1cm}

\noindent{\small{
FIG. 5. The equation of state, pressure $P$ vs. energy density $\varepsilon$ 
in the GM1 set. The results are for the nucleon-only ($np$) (proto)neutron star 
matter (top panel) and with further inclusion of $K^-$ and $\bar K^0$ 
condensation (bottom panel) calculated with the antikaon optical potential 
depth at normal nuclear matter 
density of $U_{\bar K}= -160$ MeV. The equation of state for the neutrino free 
matter is denoted by the solid line and that of the neutrino trapped matter 
by the dashed line. }} 

 \newpage
\vspace{-2cm}

{\centerline{
\epsfxsize=20cm
\epsfysize=22cm
\epsffile{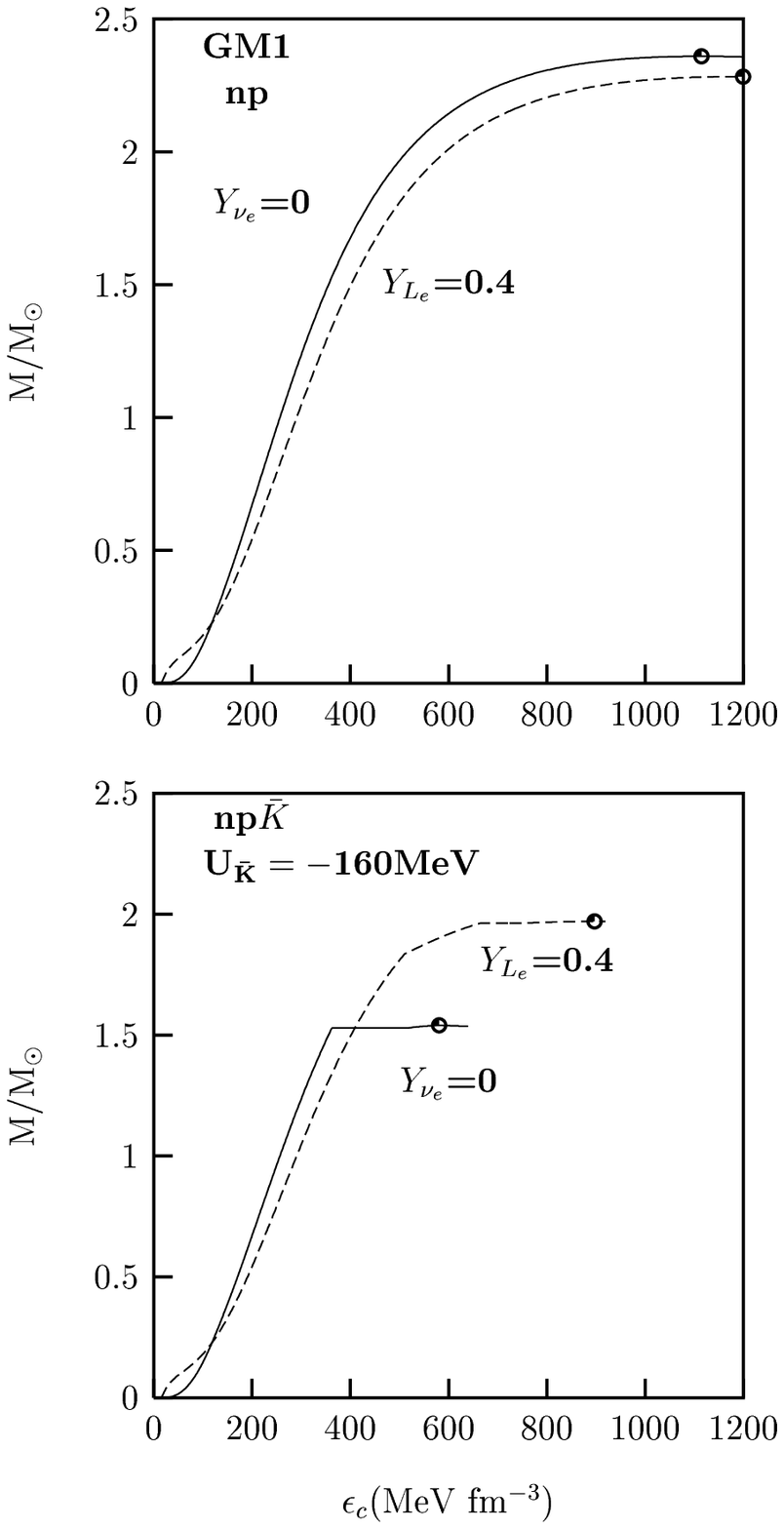}
}}

\vspace{-3cm}

\noindent{\small{
FIG. 6. The (proto)neutron star mass sequences are plotted with central energy 
density in the GM1 set for the antikaon optical potential depth 
of  $U_{\bar K}=-160$ MeV. The star masses of the nucleon-only matter and
with further inclusion of $K^-$ and $\bar K^0$ condensation are shown  
in the top and bottom panel, respectively. The solid curve corresponds to the 
neutrino free case whereas the dashed curve represents the neutrino trapped 
case. The filled circles correspond to the maximum masses.}}
 \newpage
\vspace{-2cm}

{\centerline{
\epsfxsize=14cm
\epsfysize=17cm
\epsffile{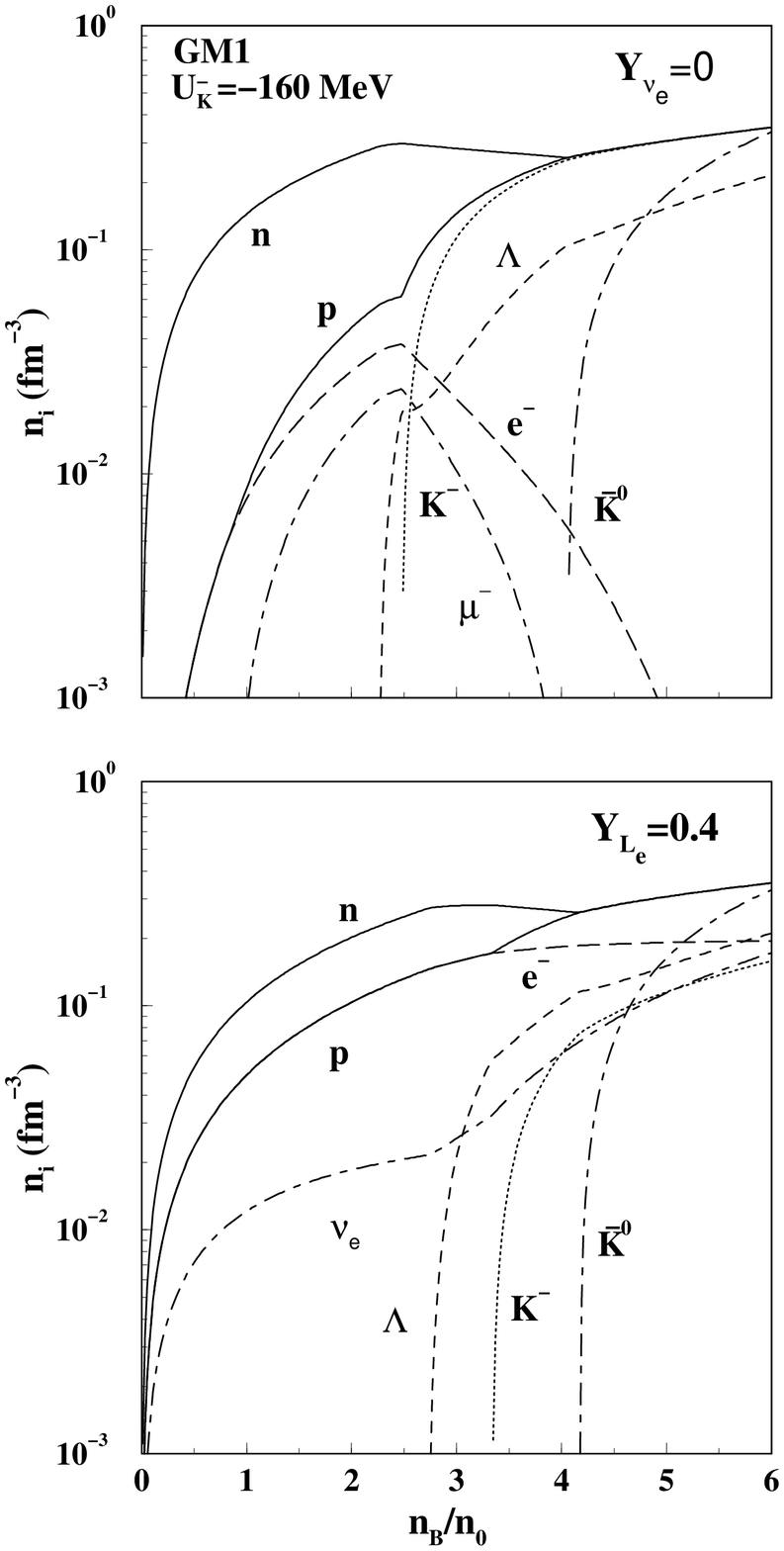}
}}

\vspace{1cm}

\noindent{\small{
FIG. 7. Same as FIG. 4, but for hyperon matter including $K^-$ and 
$\bar K^0$ condensation in the GM1 set. The results are for the neutrino 
free case (top panel) and the neutrino trapped case (bottom panel).}}
 \newpage
\vspace{-2cm}

{\centerline{
\epsfxsize=20cm
\epsfysize=22cm
\epsffile{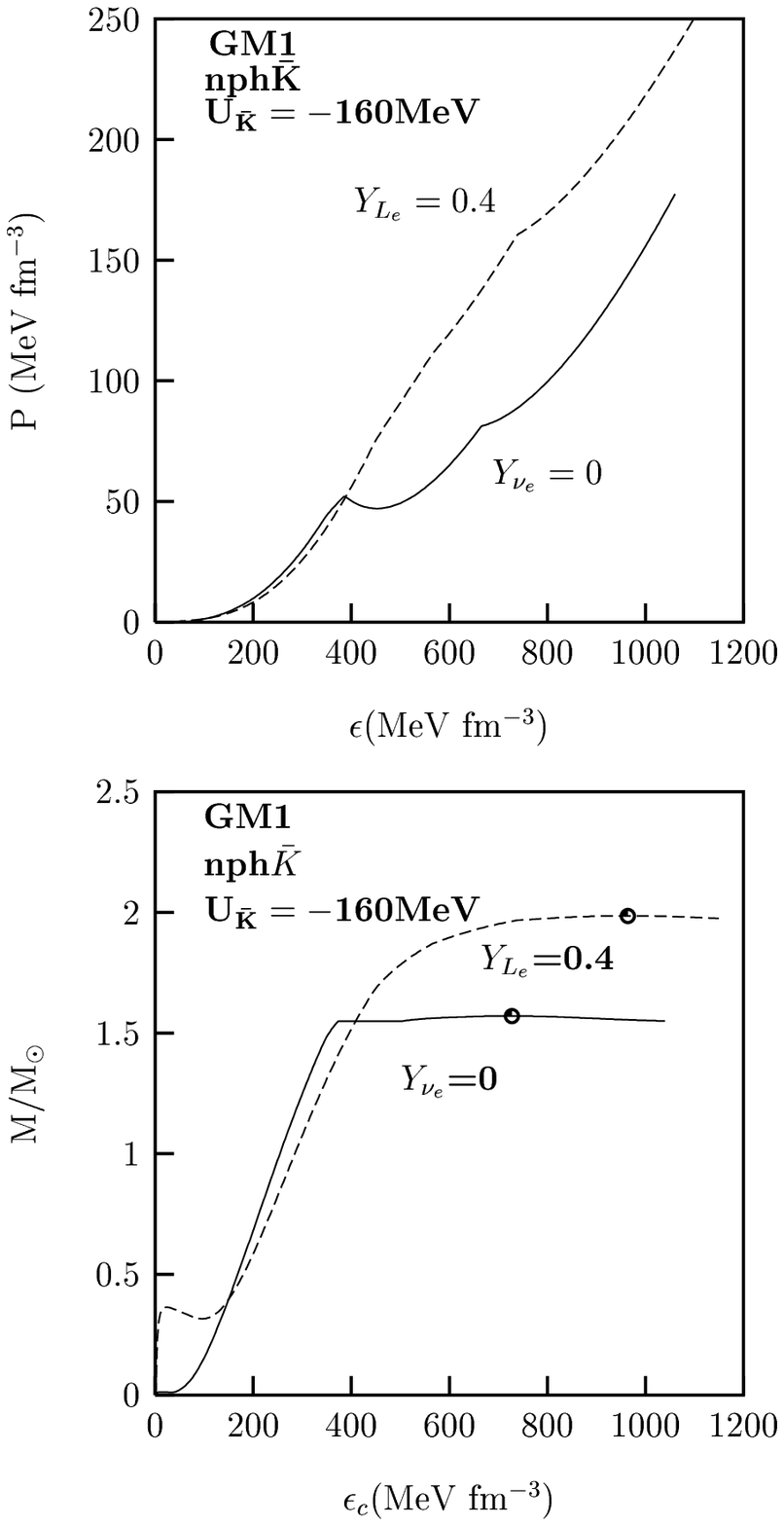}
}}

\vspace{-3cm}

\noindent{\small{
FIG. 8. The equations of state (top panel) for the neutrino free(trapped) 
hyperon 
matter including antikaon condensates with energy density in the GM1 set. The
mass sequences (bottom panel) of the (proto)neutrino stars including antikaon 
condensates are shown with central energy density in the GM1 set. The 
calculations are performed for $U_{\bar K}(n_0)=-160$ MeV. The solid and 
dashed lines correspond to the neutrino free and trapped case, respectively.
The filled circles correspond to the maximum masses.}}
 
\end{document}